\documentclass[prl,twocolumn,showpacs]{revtex4}

\usepackage{amsmath}%
\usepackage{graphicx}
\usepackage{dcolumn}
\usepackage{bm}

\begin{document}

\preprint{ITP/UU-XXX}

\title{Renormalization Group Theory for the Imbalanced Fermi Gas}

\author{K. B. Gubbels}
\email{K.Gubbels@phys.uu.nl}
\author{H. T. C. Stoof}

\affiliation{
Institute for Theoretical Physics, Utrecht University,\\
Leuvenlaan 4, 3584 CE Utrecht, The Netherlands}


\begin{abstract}
We formulate a wilsonian renormalization group theory for the
imbalanced Fermi gas. The theory is able to recover quantitatively
well-established results in both the weak-coupling and the
strong-coupling (unitarity) limit. We determine for the latter
case the line of second-order phase transitions of the imbalanced
Fermi gas and in particular the location of the tricritical point.
We obtain good agreement with the recent experiments of Y.~Shin
{\it et al}. [Nature {\bf 451}, 689 (2008)].
\end{abstract}


\maketitle

\emph{Introduction.} --- The amazing experimental control in the
manipulation of degenerate Fermi mixtures, has led for the
balanced mixture to an accurate study of the crossover between a
Bardeen-Cooper-Schrieffer (BCS) superfluid and a Bose-Einstein
condensate (BEC) of diatomic molecules. Particularly interesting
is the strongly-interacting regime, where the scattering length of
the interaction becomes much larger than the average interatomic
distance. In this so-called unitarity limit, experiments have
revealed that the superfluid state is remarkably stable and has a
record-high critical temperature of about one tenth of the Fermi
energy \cite{Jin}. Theoretically, the unitarity limit is extremely
challenging, because there is no rigorous basis for perturbation
theory due to the lack of a small parameter. As a result,
mean-field theory is only useful for understanding the relevant
physics qualitatively, but cannot be trusted quantitatively. In
order to get accurate results, more sophisticated theoretical
methods have to be invoked.

An important example is using quantum Monte-Carlo techniques,
which can provide exact results about the strongly-interacting
regime \cite{Carlson,Astrakharchik,Burovski,Lobo}, but offer less
physical insight than analytic methods. Therefore, several other
approaches have been developed to improve on mean-field theory.
Examples are theories incorporating Gaussian fluctuations
\cite{Melo,Haussmann,Strinati,Parish,Hu}, $\epsilon$ expansion
\cite{Son}, $1/N$ expansion \cite{Veillette,Sachdev}, and the
functional renormalization group (RG) \cite{Birse,Diehl}. In this
Letter, we formulate a so-called wilsonian RG to study the
strongly-interacting atomic Fermi mixture with a population
imbalance. The intuitively appealing wilsonian approach, which has
been extremely successful in the study of critical phenomena
\cite{Wilson}, is based on systematically integrating out
short-wavelength degrees of freedom, which then renormalize the
coupling constants in the effective action for the long-wavelength
degrees of freedom. For fermions, the excitations of lowest energy
lie near the Fermi level, which is therefore the natural end point
for a renormalization group flow \cite{Shankar}. A notorious
problem for interacting fermions is that under renormalization the
Fermi level also flows to an {\it a priori} unknown value, making
the wilsonian RG difficult to perform in practice
\cite{Honerkamp,Kopietz}. We show, however, how to obtain RG
equations that automatically flow to the final value of the
renormalized Fermi level.

The unitary, two-component Fermi mixture with an unequal number of
particles in each spin state is a topic of great interest in
atomic physics, condensed matter, nuclear matter, and
astroparticle physics. The landmark atomic-physics experiments
exploring this system, performed at MIT by Zwierlein \emph{et al.}
\cite{Ketterle} and at Rice University by Partridge \emph{et al.}
\cite{Hulet}, induced a large amount of activity, caused by an
intriguing mix of mutual consistent and contradictory results. In
summary, both experiments observed no oscillating order parameter,
so that the Fulde-Ferrell and Larkin-Ovchinnikov phases do not
seem to play a role in the unitarity limit. Therefore, the
experiments are consistent with a phase diagram including both
second-order and first-order phase transitions between the
superfluid (BCS or Sarma) phase and the normal phase, that are
connected by a tricritical point \cite{Parish,Gubbels}. However,
as a function of population imbalance Zwierlein \emph{et al.}
obtain a critical imbalance at which the trapped Fermi gas becomes
fully normal, whereas Partridge \emph{et al.} observe a superfluid
core up to their highest imbalances. Although this contradictory
result is still not completely understood, more recent work
implies that the data of Zwierlein \emph{et al.} is consistent
with the local-density approximation, whereas the experiments of
Partridge \emph{et al.} explore physics beyond this approximation,
possibly due to the smaller number of particles and the more
extreme aspect ratio of the trap \cite{Partridge,Haque}.

Since the validity of the local-density approximation implies that
the Fermi mixture can be seen as being locally homogeneous, the
MIT group is in the unique position to experimentally map out the
homogeneous phase diagram by performing local measurements in the
trap. Most recently, this important experiment was performed by
Shin \emph{et al.} \cite{Shin}, obtaining for the homogeneous
tricritical point in the unitarity limit $P_{\rm c3} = 0.20(5)$
and $T_{\rm c3} = 0.07(2)$ $T_{{\rm F}\uparrow}$, with $P$ the
local polarization given by
$P=(n_{\uparrow}-n_{\downarrow})/(n_{\uparrow}+n_{\downarrow})$,
$n_{\sigma}$ the density of atoms in spin state $|\sigma\rangle$,
$T$ the temperature, and $\epsilon_{{\rm F}\sigma}=k_{\rm
B}T_{{\rm F}\sigma}=(6\pi^2 n_{\sigma})^{2/3}\hbar^2/2 m$ the
Fermi energies with $m$ the atomic mass. So far, there has not
been an accurate calculation for this homogeneous tricritical
point. In this Letter, we determine it to lie at $P_{\rm c3}=0.24$
and $T_{\rm c3} = 0.06$ $ T_{{\rm F}\uparrow}$, in good agreement
with the experiment by Shin \emph{et al.}.

\emph{Wilsonian renormalization.} --- The central idea of
wilsonian renormalization is to subsequently integrate out degrees
of freedom in shells at high momenta $\Lambda$ of infinitesimal
width $d\Lambda$ and absorb the result of the integrations into
various coupling constants, which are therefore said to flow.
First, we calculate the Feynman diagrams renormalizing the
coupling constants of interest, while keeping the integration over
the internal momenta restricted to the considered high-momentum
shell. Only one-loop diagrams contribute to the flow, because the
thickness of the momentum shell is infinitesimal and each loop
introduces a factor $d\Lambda$. In order to obtain the exact
partition sum, it is then needed to consider an infinite number of
coupling constants. Although this is not possible in practice, the
RG is still able to distinguish between the relevance of the
various coupling constants, such that a carefully selected set of
them already leads to highly accurate results.

Consider the action of an interacting Fermi mixture
\begin{eqnarray}
&&S[\phi^*,\phi]=\sum_{{\bf k},n,\sigma}\phi^*_{\sigma{\bf
k},n}(-i\hbar\omega_n+\epsilon_{\bf
k}-\mu_{\sigma})\phi_{\sigma{\bf k},n} \\
&&+\frac{1}{\hbar\beta V}\sum_{\substack{{\bf k},{\bf k'},{\bf
q}\\{n},{n'},{m}}} \Gamma_{{\bf q},m}\phi^*_{\uparrow{\bf
q-k'},m-n'}\phi^*_{\downarrow{\bf k'},n'}\phi_{\downarrow{\bf
q-k},m-n}\phi_{\uparrow{\bf k},n}~, \nonumber
\end{eqnarray}
with $\omega_n$ the odd fermionic Matsubara frequencies,
$\epsilon_{\bf k}=\hbar^2k^2/2 m$ the kinetic energy, $\mu_\sigma$
the chemical potentials, $\beta=1/k_{\rm B}T$, $V$ the volume,
$\Gamma_{{\bf q},m}$ the interaction vertex and $\phi_{\sigma{\bf
k},n}$ the fermionic fields corresponding to annihilation of a
particle with spin $\sigma$, momentum ${\bf k}$ and frequency
$\omega_n$. In Fig.\ 1, we have drawn the Feynman diagrams
renormalizing $\mu_\sigma$ and $\Gamma_{{\bf q},m}$. To start with
a simple wilsonian RG, we take the interaction vertex to be
frequency and momentum independent. If we then consider only the
three coupling constants $\mu_{\sigma}$ and $\Gamma_{{\bf 0},0}$,
we find
\begin{eqnarray}
\frac{d\Gamma^{-1}_{{\bf 0},0}}{d\Lambda} &=&\frac{\Lambda^2}{2
\pi^2}
\left[\frac{1-N_{\uparrow}-N_{\downarrow}}{2(\epsilon_{\Lambda}-\mu)}-\frac{N_{\uparrow}-N_{\downarrow}}{2
h}\right], \label{RGI}\\
\frac{d\mu_{\sigma}}{d\Lambda}&=&-\frac{\Lambda^2}{2 \pi^2}
\frac{N_{-\sigma}}{\Gamma^{-1}_{{\bf 0},0}}~, \label{RGC}
\end{eqnarray}
with $\mu=(\mu_{\uparrow}+\mu_{\downarrow})/2$,
$h=(\mu_{\uparrow}-\mu_{\downarrow})/2$ and the Fermi distribution
$N_{\sigma}=1/\{\exp[\beta
(\epsilon_{\Lambda}-\mu_{\sigma})]+1\}$. These expressions are
readily obtained from the diagrams in Fig. 1 by setting all
external frequencies and momenta equal to zero and by performing
in each loop the full Matsubara sum over internal frequencies,
while integrating the internal momenta over the infinitesimal
shell $d\Lambda$. The first term in Eq. (\ref{RGI}) corresponds to
the ladder diagram and describes the scattering between particles.
The second term corresponds to the bubble diagram and describes
screening of the interaction by particle-hole excitations. Also
note that due to the coupling of the differential equations for
$\mu_{\sigma}$ and $\Gamma^{-1}_{{\bf 0},0}$, we automatically
generate an infinite number of Feynman diagrams, showing the
nonperturbative nature of the RG.

\begin{figure}
\includegraphics[width=1.0\columnwidth]{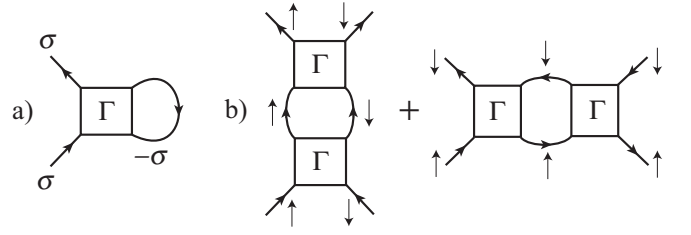}
\caption{\label{FeynDiag} Feynman diagrams renormalizing a) the
chemical potentials and b) the interatomic interaction.}
\end{figure}

However, when the Fermi mixture is critical, the inverse many-body
vertex $\Gamma^{-1}_{{\bf 0},0}$ flows to zero according to the
Thouless criterion and the chemical potentials in Eq. (\ref{RGC})
diverge, which is unphysical. To go beyond this simple RG and
calculate critical properties realistically, we need to take the
frequency and momentum dependence of the interaction vertex into
account, which are generated by the ladder and the bubble
diagrams. The ladder diagram depends only on the external
center-of-mass coordinates ${\bf q} $ and $\omega_m$, and its
contribution to the renormalization of $\Gamma^{-1}_{{\bf q},m}$
is given by
\begin{equation}\label{ladder}
\Xi(q^2,i \omega_m)=\int_{d\Lambda}d{\bf q'
}\frac{1-N_{\uparrow}(\epsilon_{\bf
q'})-N_{\downarrow}(\epsilon_{\bf
q-q'})}{-i\hbar\omega_m+\epsilon_{\bf q'}+\epsilon_{\bf
q-q'}-2\mu}~,
\end{equation}
where during integration both ${\bf q'}$ and ${\bf q-q'}$ have to
remain in the infinitesimal shell $d\Lambda$. Since the ladder
diagram is already present at the two-body level, it is most
important for scattering properties. Therefore, the interaction
vertex is mainly dependent on the center-of-mass coordinates and
we neglect the dependence of the vertex on other frequencies and
momenta. The way to treat the external frequency and momentum
dependence in wilsonian RG, is by expanding the (inverse)
interaction in the following way: $\Gamma^{-1}_{{\bf
q},m}=\Gamma^{-1}_{{\bf
0},0}-Z^{-1}_{q}q^2+Z^{-1}_{\omega}i\hbar\omega_m$. The flow
equations for the additional coupling constants $Z_{q}^{-1}$ and
$Z_{\omega}^{-1}$ are then obtained by considering the derivatives
$\partial_{q^2}\Xi(q^2,\omega)|_{q=\omega=0}$ and
$\partial_{\omega}\Xi(q^2,\omega)|_{q=\omega=0}$.

\emph{Extreme imbalance.}--- First, we apply the RG to one
spin-down particle in a Fermi sea of spin-up particles at zero
temperature in the unitarity limit. The full equation of state for
the normal state of a strongly-interacting Fermi mixture was
obtained at zero temperature using Monte-Carlo techniques
\cite{Lobo}. The most important feature of this equation is a
so-called mean-field shift, caused by the strong interactions and
characterized by a parameter $A$, which describes the self-energy
of a single spin-down particle in a sea of spin-up particles
\cite{Lobo,Combescot}. In this case, the RG equations are
simplified, because $N_{\downarrow}$ can be set to zero and thus
$\mu_{\uparrow}$ is not renormalized. Next, we have to incorporate
the momentum and frequency dependence of the interaction in the
one-loop Feynman diagram for the renormalization of
$\mu_{\downarrow}$. In this particular case, the external
frequency dependence of the ladder diagram can be taken into
account exactly, since the one-loop Matsubara sum simply leads to
the substitution $i\hbar \omega_m\rightarrow \epsilon_{\bf
q}-\mu_{\uparrow}$ in Eq. (\ref{ladder}) \cite{Combescot}. The
external momentum dependence is accounted for by the coupling
$Z_{q}^{-1}$, giving
\begin{eqnarray}
\frac{d\Gamma^{-1}_{{\bf 0},0}}{d\Lambda} &=&\frac{\Lambda^2}{2
\pi^2}
\left[\frac{1-N_{\uparrow}}{2\epsilon_{\Lambda}-\mu_{\downarrow}}-\frac{N_{\uparrow}}{2
h}\right]~,\\
\frac{d\mu_{\downarrow}}{d\Lambda}&=&\frac{\Lambda^2}{2 \pi^2}
\frac{N_{\uparrow}}{-\Gamma^{-1}_{{\bf 0},0}+Z^{-1}_{q}\Lambda^2}~,\\
\frac{dZ^{-1}_{q}}{d\Lambda} &=& - \frac{\hbar^4\Lambda^4}{6
\pi^2m^2}
\frac{1-N_{\uparrow}}{(2\epsilon_{\Lambda}-\mu_{\downarrow})^3}~,
\end{eqnarray}
where we note that these equations only have poles for positive
values of $\mu_{\downarrow}$. Since this will not occur, we can
simply use $\Lambda(l)=\Lambda_0 e^{-l}$ to integrate out all
momentum shells \cite{Bijlsma}. We then obtain a system of three
coupled ordinary differential equations in $l$, which are very
easily solved numerically. If we take as an initial condition
$\Gamma^{-1}_{{\bf 0},0}(0)=-m(\pi+2|a|\Lambda_0)/4
\pi^2|a|\hbar^2$ for a negative scattering length $a$, we
automatically incorporate the relevant two-body physics exactly
into our theory and also eliminate all dependence on the
high-momentum cut-off $\Lambda_0$ \cite{Bijlsma}. The unitarity
limit is then given by $\Gamma^{-1}_{{\bf 0},0}(0)=-m
\Lambda_0/2\pi^2\hbar^2$. The other initial conditions are
$\mu_{\downarrow}(0)=\mu_{\downarrow}$ and $Z_{q}^{-1}(0)=0$,
since the interaction starts out as being momentum independent.
Note that in this calculation
$\mu_{\downarrow}(0)=\mu_{\downarrow}$ is indeed negative and
increases during the flow due to the strong attractive
interactions. The quantum phase transition from a zero density to
a nonzero density of spin-down particles occurs for the initial
value $\mu_{\downarrow}$ that at the end of the flow precisely
leads to $\mu_\downarrow(\infty)=0$. This happens when
$\mu_{\downarrow}=-0.598 \mu_{\uparrow}$, yielding $A=0.997$ in
very good agreement with the Monte Carlo result $A=0.97(2)$
\cite{Lobo}. This calculation also shows that it is crucial to let
the chemical potential flow.

\emph{Phase diagram.}--- Next, we turn to our main topic, namely
the critical properties of the strongly-interacting Fermi mixture
and the calculation of the tricritical point in the phase diagram.
Since it is not exact to make the substitution $\hbar \omega
\rightarrow \epsilon_{\bf q} - \mu_{-\sigma}$ at nonzero
temperatures, we take the frequency dependence of the ladder
diagram into account through the renormalization of the coupling
$Z^{-1}_{\omega}$. While the flow of $\Gamma^{-1}_{{\bf 0},0}$ is
still given by Eq. (\ref{RGI}), the expressions for the flow of
$\mu_{\sigma}$ and $Z_{\omega}^{-1}$ become
\begin{eqnarray}
\frac{d\mu_{\sigma}}{d\Lambda}&=&\frac{\Lambda^2}{2 \pi^2}
\frac{N_{-\sigma}+N_{\rm B}}{-\Gamma^{-1}_{{\bf
0},0}+Z^{-1}_{q}\Lambda^2 -Z^{-1}_{\omega
}(\epsilon_{\Lambda}-\mu_{-\sigma})}~, \label{RGCI}\\
\frac{dZ^{-1}_{\omega}}{d\Lambda} &=&\frac{\Lambda^2}{2 \pi^2}
\frac{1-N_{\uparrow}-N_{\downarrow}}{4(\epsilon_{\Lambda}-\mu)^2}~,
\end{eqnarray}
with $N_{\rm B}=1/\{\exp[\beta Z_{\omega}(-\Gamma^{-1}_{{\bf 0
},0}+Z^{-1}_{q}\Lambda^2)]-1\}$ coming from the bosonic frequency
dependence of the interaction. A more cumbersome expression holds
for $Z^{-1}_{q}$. The initial conditions are the same as for the
extremely imbalanced case with in addition
$\mu_{\uparrow}(0)=\mu_{\uparrow}$ and $Z^{-1}_{\omega}(0)=0$. As
mentioned before, the critical condition is that the fully
renormalized vertex $\Gamma^{-1}_{{\bf 0},0}(\infty)$, which can
be seen as the inverse many-body T-matrix at zero external
momentum and frequency, goes to zero. Physically, this implies
that a many-body bound-state is entering the system. From Eq.\
(\ref{RGCI}), we see that incorporating the coupling constants
$Z^{-1}_{q}$ and $Z^{-1}_{\omega}$, thereby taking the dependence
of the interaction on the center-of-mass momentum and frequency
into account, is crucial to solve the previously mentioned problem
of the diverging chemical potential.

The only pole left in our set of RG equations is the average Fermi
level $\mu=(\mu_{\uparrow}+\mu_{\downarrow})/2$, which is
therefore the natural end point of our RG. However, this Fermi
level is shifting due to the renormalization of the individual
chemical potentials. This problem is conveniently solved by
integrating out all momentum shells with the following procedure.
First, we start at a high momentum cutoff $\Lambda_0$ and flow to
a momentum $\Lambda'_0$ at roughly two times the average Fermi
momentum, with $k_{{\rm F}\sigma}=\sqrt{2m\epsilon_{{\rm
F}\sigma}}/\hbar$. This integrates out the high-energy two-body
physics, but hardly affects the chemical potentials. Then, we
start integrating out the rest of the momentum shells
symmetrically with respect to the flowing average Fermi level.
This is achieved by using $\hbar\Lambda_{+}(l)/\sqrt{2m}= (\hbar
\Lambda'_0/\sqrt{2m}- \sqrt{\mu})e^{-l}+\sqrt{\mu(l)}$ and by
$\hbar\Lambda_{-}(l)/\sqrt{2m}= - \sqrt{\mu}e^{-l}+\sqrt{\mu(l)}$.
Note that as desired $\Lambda_{+}(l)$ starts at $\Lambda'_0$ and
automatically flows from above to $\sqrt{2 m\mu(\infty)}/\hbar$,
whereas $\Lambda_{-}(l)$ starts at 0 and automatically flows from
below to $\sqrt{2 m\mu(\infty)}/\hbar$.

We first apply the above procedure to study the equal density
case, i.e., $h\rightarrow 0$, as a function of negative scattering
length $a$. The scattering length enters the calculation through
the initial condition of $\Gamma^{-1}_{{\bf 0},0}$. To express our
results in terms of the Fermi energy $\epsilon_{\rm F}=
\epsilon_{{\rm F}\sigma}$, we calculate the densities of atoms
with the flow equation $d n_{\sigma}/d\Lambda= \Lambda^2
N_{\sigma}/2\pi^2$. In the weak-coupling limit, $a \rightarrow
0^{-}$, the chemical potentials hardly renormalize, so that only
Eq.\ (\ref{RGI}) is relevant. The critical temperature becomes
exponentially small, which allows us to solve Eq.\ (\ref{RGI})
exactly with the result $k_{\rm B}T_{\rm c}=8\epsilon_{\rm
F}e^{\gamma-3}\exp\{-\pi/2k_{\rm F} |a|\}/\pi$ and $\gamma$
Euler's constant. Compared to the standard BCS-result we have an
extra factor of $1/e$, coming from the screening effect of the
bubble diagram that is not present in BCS theory. It is to be
compared with the so-called Gor'kov correction, that reduces the
critical temperature by a factor of 2.2 in the weak-coupling
BCS-limit \cite{Heiselberg}. The difference is due to the fact
that we have only allowed for a nonzero center-of-mass momentum.
This approximation is actually most appropriate in the unitarity
limit and expected to be less accurate for weak coupling.

\begin{figure}
\includegraphics[width=0.9\columnwidth]{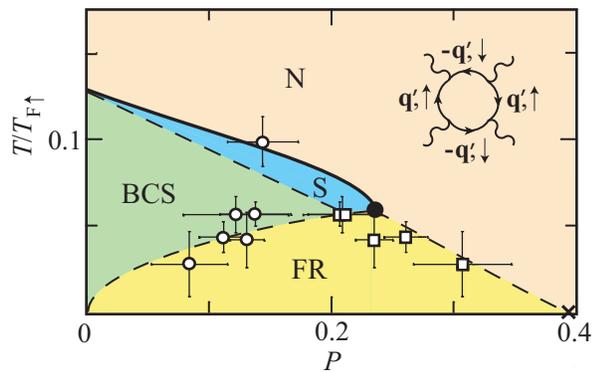}
\caption{\label{HomPhasDiag} (Color online) The phase diagram of
the homogeneous two-component Fermi mixture in the unitarity
limit, consisting of the superfluid Sarma (S) and BCS phases, the
normal phase (N) and the forbidden region (FR). The solid black
line is the result of our RG calculations. The Monte-Carlo result
of Lobo et al. \cite{Lobo}, which is recovered by our RG, is
indicated by a cross. The open circles and squares are data along
the phase boundaries from the experiment of Shin et al.
\cite{Shin}. The dashed lines are only guides to the eye. Also
shown is the Feynman diagram determining the tricritical point.}
\end{figure}

At larger values of $|a|$, the flow of the chemical potential
becomes important and we obtain higher critical temperatures. In
the unitarity limit, when $a$ diverges, we obtain $T_{\rm c}=0.13
T_{\rm F}$ and $\mu (T_{\rm c}) = 0.55 \epsilon_{\rm F}$ in good
agreement with the Monte-Carlo results $T_{\rm c}=0.152(7) T_{\rm
F}$ and $\mu(T_{\rm c}) = 0.493(14) \epsilon_{\rm F}$
\cite{Burovski}. With our RG approach, we are in the unique
position to accurately calculate the critical temperature as a
function of polarization $P$ and compare with the recent
experiment of Shin {\it et al}.. The result is shown in Fig. 2.
The inset of this figure shows the one-loop diagram determining
the position of the tricritical point. If it changes sign, then
the fourth-order coefficient in the Landau theory for the
superfluid phase transition changes sign and the nature of the
phase transition changes from second order to first order. This
yields $P_{\rm c3}=0.24$ and $T_{\rm c3} = 0.06$ $ T_{{\rm
F}\uparrow}$ in good agreement with the experimental data. Our
previous confirmation of the Monte-Carlo equation of state at
$T=0$ implies that we also agree with the prediction of a quantum
phase transition from the superfluid to the normal phase at a
critical imbalance of $P_{\rm c}=0.39$ \cite{Lobo}. Note that up
to now, all theoretical predictions for the location of the
tricritical point do not fit on the scale of Fig. 2. However, our
calculations find good agreement with the experiments of Shin
\emph{et al}. in all limits.

Near the second-order phase boundary, the BCS order parameter
$|\Delta|$ becomes arbitrarily small. Since at nonzero
polarization we have that $h(\infty)>0$, it immediately follows
that $h(\infty)>|\Delta|$. This means that the normal gas is
unstable towards the so-called Sarma phase, which is a polarized
superfluid with a gapless excitation spectrum for the majority
spin-species \cite{Gubbels}. However, the present RG is not
suitable to calculate the full extent of the Sarma phase in the
phase diagram or the precise shape of the forbidden region,
because this requires a RG for the superfluid phase. The
corresponding calculations are more involved than the present RG
for the normal phase and is work in progress.

\emph{Acknowledgements.}
--- We thank Tilman Enss and Pietro Massignan for useful discussions and
Yong-il Shin for kindly providing us with the experimental data.
This work is supported by the Stichting voor Fundamenteel
Onderzoek der Materie (FOM) and the Nederlandse Organisatie voor
Wetenschaplijk Onderzoek (NWO).

\end{document}